\documentclass{emulateapj}

\slugcomment{Draft version, \today}

\shorttitle{Dynamical State of XMMXCS J$2215.9-1738$}
\shortauthors{Hilton et al.}

\begin{document}

%=============================================================================================
\title{The \textit{XMM} Cluster Survey: The Dynamical\\ State of XMMXCS J$2215.9-1738$ at \lowercase{$z=1.457$}}

%=============================================================================================
\author{Matt Hilton\altaffilmark{1}, Chris A. Collins\altaffilmark{1}, S. Adam Stanford\altaffilmark{2, 3}, Chris Lidman\altaffilmark{4}, Kyle S. Dawson\altaffilmark{5}\\Michael Davidson\altaffilmark{6}, Scott T. Kay\altaffilmark{7}, Andrew R. Liddle\altaffilmark{8}, Robert G. Mann\altaffilmark{6}, Christopher J. Miller\altaffilmark{9}\\Robert C. Nichol\altaffilmark{10}, A. Kathy Romer\altaffilmark{8}, Kivanc Sabirli\altaffilmark{11}, Pedro T. P. Viana\altaffilmark{12,13}, Michael J. West\altaffilmark{14}}

%=============================================================================================
\altaffiltext{1}{Astrophysics Research Institute, Liverpool John Moores University, Twelve Quays House, Egerton Wharf, Birkenhead, CH41 1LD, UK; mjh@astro.livjm.ac.uk}
\altaffiltext{2}{University of California, Davis, CA 95616, USA; adam@igpp.ucllnl.org}
\altaffiltext{3}{Institute of Geophysics and Planetary Physics, Lawrence Livermore National Laboratory, Livermore, CA 94551, USA}
\altaffiltext{4}{European Southern Observatory, Vitacura, Santiago 19, Chile}
\altaffiltext{5}{E. O. Lawrence Berkeley National Laboratory, Berkeley, CA 94720, USA}
\altaffiltext{6}{Institute of Astronomy, University of Edinburgh, Blackford Hill, Edinburgh, EH9 9HJ, UK}
\altaffiltext{7}{University of Manchester, Jodrell Bank Observatory, Macclesfield, Cheshire, SK11 9DL, UK}
\altaffiltext{8}{Astronomy Centre, University of Sussex, Falmer, Brighton, BN1 9QH, UK}
\altaffiltext{9}{Cerro-Tololo Inter-American Observatory, National Optical Astronomy Observatory, 950 North Cherry Ave., Tucson, AZ 85719, USA}
\altaffiltext{10}{Institute of Cosmology and Gravitation, University of Portsmouth, Portsmouth, PO1 2EG, UK}
\altaffiltext{11}{Department of Physics, Carnegie Mellon University, 5000 Forbes Ave., Pittsburgh, PA 15217, USA}
\altaffiltext{12}{Departmento de Matem\'{a}tica Aplicada da Faculdade de Ci\^{e}ncias da Universidade do Porto, Rua do Campo Alegre, 687, 4169-007, Portugal}
\altaffiltext{13}{Centro de Astrof\'{\i}sica da Universidade do Porto, Rua das Estrelas, 4150-762, Porto, Portugal}
\altaffiltext{14}{Gemini Observatory, Casilla 603, La Serena, Chile}

%=============================================================================================
\begin{abstract}
We present new spectroscopic observations of the most distant X-ray selected galaxy cluster currently known, XMMXCS J2215.9$-$1738 at $z=1.457$, obtained with the DEIMOS instrument at the W. M. Keck Observatory, and the FORS2 instrument on the ESO Very Large Telescope. Within the cluster virial radius, as estimated from the cluster X-ray properties, we increase the number of known spectroscopic cluster members to 17 objects, and calculate the line of sight velocity dispersion of the cluster to be $580 \pm 140$ km s$^{-1}$. We find mild evidence that the velocity distribution of galaxies within the virial radius deviates from a single Gaussian. We show that the properties of J2215.9$-$1738 are inconsistent with self-similar evolution of local X-ray scaling relations, finding that the cluster is underluminous given its X-ray temperature, and that the intracluster medium contains $\sim 2-3$ times the kinetic energy per unit mass of the cluster galaxies. These results can perhaps be explained if the cluster is observed in the aftermath of an off-axis merger. Alternatively, heating of the intracluster medium through supernovae and/or Active Galactic Nuclei activity, as is required to explain the observed slope of the local X-ray luminosity--temperature relation, may be responsible.
\end{abstract}

%=============================================================================================
\keywords{X-rays: galaxies: clusters --- galaxies: clusters: individual (XMMXCS J2215.9$-$1738) }

%=============================================================================================
\section{Introduction}
\label{s_SpectroIntro}
Galaxy clusters are both important cosmological probes and places in which to study the effects of environment upon galaxy evolution. As the most massive gravitationally bound structures in the universe, clusters are the last objects to form in the hierarchy of structure formation, and their abundance and spatial distribution is therefore extremely sensitive to the matter density of the universe \citep[see][for a recent review]{Voit_2005}. Cluster catalogs constructed from surveys at X-ray wavelengths provide a powerful tool with which to test cosmological models, because X-ray observables, such as luminosity and temperature, are readily related to cluster mass. X-ray surveys conducted with the \textit{ROSAT} satellite showed that the universe is low density ($\Omega_m \approx 0.3$), from the observed spatial distribution of clusters \citep{Collins_2000}, and the cluster mass function \citep{Borgani_2001, Reiprich_2002, Schuecker_2003}. Cluster surveys reaching to high redshift ($z \sim 1$) are able to constrain the dark energy density, through the observed evolution in the mass function with redshift (e.g. \citealt*{Carlstrom_2002}, \citealt{Mohr_2005}, \citealt{RomerXCS_2001}). Studies of the gas mass fraction within clusters can also be used to measure the dark energy density \citep[e.g.][]{Allen_2004}. It is important to note that these constraints are independent of those obtained from cosmic microwave background observations \citep[e.g.,][]{Spergel_2006}, or surveys using type Ia supernovae as standard candles (e.g., \citealt{Astier_2006, Riess_2007, WoodVasey_2007}, for a compilation of all supernova surveys see Kowalski et al. 2007, in preparation). 

However, to infer cosmological parameter estimates from X-ray selected cluster samples, it is essential that the redshift evolution of the mass-scaling relations with luminosity ($L_{\rm X}$) and temperature ($T$) are understood, as unaccounted for evolution in the properties of the intracluster medium (ICM) would lead to an erroneous determination of evolution in the mass function. Although it is agreed that the slope of the $L_{\rm X}-T$ relation is not consistent with the expectation of `self-similar' evolution of the ICM, where gravity is the sole process responsible for setting the observed properties \citep{Kaiser_1986}, investigations to date, using cluster samples that reach to $z \sim 1$, do not reach consensus as to the form of the evolution of this relation. Several authors \citep{Vikhlinin_2002, Lumb_2004, Maughan_2006} find evidence that the $L_{\rm X}-T$ relation evolves in a way consistent with self-similar evolution, while others suggest that the evolution is much milder, perhaps even negative \citep{Ettori_2004}. However, the large amount of intrinsic scatter in the relation means that the only way in which a firm conclusion can be drawn is by extending such studies to include the $z>1$ cluster population, as evolution predictions in models that include the effects of different non-gravitational heating processes on the ICM deviate significantly at such redshifts (\citealt*{Muanwong_2006}, \citealt{Maughan_2006}).

The search for galaxy clusters at $z >1$ is also crucial for increasing our understanding of the processes that shaped the evolution of the elliptical galaxies that dominate the cluster population, at least up to $z\sim1.3$ \citep{Blakeslee_2003}, because the epoch of cluster formation is expected to be $1 < z < 2$ in the $\Lambda$CDM cosmology. However, to date, few objects have been discovered at such high redshifts \citep{Andreon_2005,  Bremer_2006, Brodwin_2006, Hashimoto_2004, Mullis_2005, Rosati_1999, Rosati_2004, Stanford_1997, Stanford_2002, Stanford_2005}.

In this paper, we report new spectroscopic observations of XMMXCS J2215.9$-$1738 at $z=1.457$ \citep{Stanford_2006}, which was identified as an extended X-ray source in the \textit{XMM} Cluster Survey \citep[XCS\footnotemark;][]{RomerXCS_2001}\footnotetext{http://www.xcs-home.org}, an ongoing serendipitous search for galaxy clusters in the \textit{XMM-Newton} Science Archive\footnotemark (XSA)\footnotetext{http://xmm.vilspa.esa.es/xsa/}. XCS has the principal aim of measuring the cosmological parameters $\sigma_8$ (the variance of the mass density on a scale of 8 Mpc, i.e., the normalization of the dark matter power spectrum), $\Omega_m$, and $\Omega_\Lambda$. Spectroscopic follow-up of J2215.9$-$1738 with the DEep Imaging Multi Object Spectrograph \citep[DEIMOS;][]{Faber_2003} on the 10m Keck II telescope secured six galaxies with concordant redshifts at $z=1.45$ within $30 \arcsec$ of the cluster X-ray centroid. The cluster was found to have bolometric X-ray luminosity $L_{\rm X} = 4.39^{+0.46}_{-0.37} \times 10^{44}$ ergs s$^{-1}$, and temperature $kT = 7.4^{+1.6}_{-1.1}$ keV, making it the hottest cluster known at $z>1$ \citep[][note that throughout this paper all quoted X-ray luminosities are bolometric, and we quote all uncertainties as 68 per cent confidence limits]{Stanford_2006}.

The structure of this paper is as follows. In \textsection~\ref{s_SpectroObservations}, we describe observations of J2215.9$-$1738 performed at the W. M. Keck Observatory and the ESO Very Large Telescope (VLT). We describe the data reduction and redshift measurements in \textsection~\ref{s_SpectroDataReduction} and \textsection~\ref{s_SpectroAnalysis}. In \textsection~\ref{s_SpectroResults}, we present the newly identified cluster members, the cluster velocity distribution, and the measurement of the line of sight velocity dispersion. In \textsection~\ref{s_SpectroDiscuss}, we discuss the relationships between the velocity dispersion of J2215.9$-$1738 and its X-ray properties. 

We assume a concordance cosmology of $\Omega_m=0.3$, $\Omega_\Lambda=0.7$, and $H_0=70$ km s$^{-1}$ Mpc$^{-1}$ throughout, where $\Omega_\Lambda$ is the energy density associated with a cosmological constant.

%=============================================================================================
\section{Observations}
\label{s_SpectroObservations}
Spectroscopic observations were obtained using the DEIMOS instrument on Keck II, and the FOcal Reducer and low dispersion Spectrograph \citep[FORS2;][]{Appenzeller_1998} on the 8m ESO Very Large Telescope (VLT) Unit 1 (Antu). Both of these instruments are capable of conducting multi-object spectroscopy (MOS), using custom designed slit masks. The capabilities of each instrument vary significantly: DEIMOS covers a $16.7\arcmin \times 5.0\arcmin$ field of view, and can use masks with $>100$ slitlets; FORS2 has a field of view of $6.8 \arcmin \times 6.8 \arcmin$, and in MOS mode (as used for our observations) can target up to 19 objects simultaneously using a series of moveable slitlets. Both instruments have very red sensitive CCD detectors with quantum efficiency $\sim 50 $ per cent at 9500 \AA{}. 

As described in \citet{Stanford_2006}, target galaxies for spectroscopic observations were selected to lie in the appropriate region of the $K_s$, $I-K_s$ colour-magnitude diagram for the red-sequence of a $z=1.45$ cluster -- i.e., with $I>22$, $I-K>4$ (magnitudes are on the Vega system). The faintest galaxy for which a secure redshift measurement was obtained has a magnitude of $I=23.9$. We describe the observations obtained with each telescope in turn below. Table~\ref{t_spectroObsData} presents a log of the observations. 
   
%---------------------------------------------------------------------------------------------
\subsection{Keck}
For all our Keck observations, DEIMOS was used with the OG550 order sorting filter and the 600ZD grating, which is blazed at 7500 \AA{}, has dispersion of 0.65 \AA{} pixel$^{-1}$, and provides typical wavelength coverage of 5000-10000 \AA{}. Slits of width $1\arcsec$ and minimum length 5$\arcsec$ were used. In 2006, we have obtained observations using three new slit masks covering the field of J2215.9$-$1738 using DEIMOS. $5 \times 1800$ sec of exposure were obtained on behalf of the XCS team by P. Guhathakurta on UT 2006 September 16. Observations using two further masks were obtained on UT 2006 September 20 and 21, during which weather conditions were reasonable (some thin cirrus) and seeing was good (typically $0.6-0.8 \arcsec$). As the seeing was less than the slit width, the spectral resolution of the DEIMOS data varied between $3.3-5.5$ \AA{}, which is sufficient to resolve the 3727 \AA{} [O \textsc{ii}] emission line into two components at the redshift of the cluster. We obtained $8 \times 1800$ sec exposures on the 20th, and $7 \times 1800$ sec exposures on the 21st. 

%---------------------------------------------------------------------------------------------
\subsection{VLT} 
Hubble Space Telescope (HST) Advanced Camera for Surveys (ACS) observations of J2215.9$-$1738 have been obtained through a program designed to place constraints on the dark energy through observations of high redshift Type Ia supernovae\footnotemark. The program targeted several $z > 1$ clusters, because the elliptical galaxies they contain are relatively free of dust, which is one of the largest sources of systematic and statistical uncertainty in using Type Ia supernovae for cosmology. A Type Ia supernova candidate was detected in ACS imaging of J2215.9$-$1738 on UT 2006 June 19, and prompted follow-up spectroscopic observations by the supernova survey team using the FORS2 instrument on VLT Unit 1. 
\footnotetext{Based on observations made with the NASA/ESA Hubble Space Telescope, obtained from the data archive at the Space Telescope Institute. STScI is operated by the association of Universities for Research in Astronomy, Inc. under the NASA contract NAS 5-26555. The observations are associated with program 10496.}

For the VLT observations, FORS2 was used with the 300I grism and the OG590 order sorting filter. This configuration has a dispersion of 3.2\AA{} pixel$^{-1}$ and provides a wavelength range starting from 5900\AA{} and extending to approximately 10000\AA{}. Since the observations had to be carried out at short notice (the SN had to be observed before it faded from view), the observations were done with the MOS mode of FORS2. The MOS mode consists of 19 moveable slits with lengths that vary between $20-22\arcsec$. The slit width was set to 1$\arcsec$.  All the data were taken during clear nights and in $0.6-1.1\arcsec$ seeing. Since the seeing was often narrower than the slit width, the resolution of the FORS2 spectra varies between $7.7-12.8$ \AA{}.

The field of J2215.9$-$1738 was observed with four different MOS configurations. The first configuration was used when the SN was near maximum light. The other three configurations were done several months later when the SN was significantly fainter. In all masks, a slit was placed on the SN and the host galaxy, thus a spectrum of the SN with the host and a spectrum of the host alone was obtained. The other slits were placed on candidate cluster members or field galaxies. For each MOS set-up, between three to nine 900 second exposures were taken. Between each exposure, the telescope was moved a few arcseconds along the slit direction. These offsets, which shift the spectra along detector columns, were done for two reasons: firstly, they minimize the possibility that an object is lost because its spectrum lands on a bad row in one of the detectors; and secondly, the data can then be used in fringe removal, the process of which is described in further detail in \textsection~\ref{s_VLTReduction}.

A total of 47 slits across four MOS masks were used to observe 28 independent targets, one of which was an alignment star that was common to all masks. Some target candidate cluster members were also observed in more than one mask. Note that these duplicate observations were not used to infer the uncertainty in the subsequently described redshift measurements: the target objects in question were simply faint. From 27 slits (excluding the one that was placed on the star), 41 redshifts were obtained. The reason for the high redshift efficiency is that the MOS slits are relatively long and there is often more that one target in the slit, resulting in a high number of serendipitous redshifts.

%=============================================================================================
\section{Data reduction}
\label{s_SpectroDataReduction}
\subsection{Keck}
The DEIMOS data were reduced using version v1.1.4 of \texttt{spec2d}, the automated data reduction pipeline developed by the DEEP2 galaxy redshift survey team \citep{Davis_2003}. Firstly, the flat and wavelength calibration frames are processed and the location of slitlets identified. A 2D wavelength solution is computed for each slitlet using the wavelength calibration images and the DEIMOS optical model. We found that the wavelength calibration is accurate to approximately 0.07 \AA{}, from a comparison of the locations of several bright OH sky emission lines in a subset of the object spectra with the tables of \citet{OsterbrockMartel_1992}. At the next stage the science frames are processed: the data are bias subtracted, flat-fielded, and a spline model of the sky spectrum in each slitlet is constructed. No correction is made for fringing effects at red wavelengths, as this is not a significant feature of spectra obtained with DEIMOS. The science frames are then combined to produce a mean, sky subtracted 2D spectrum cleaned of cosmic rays for each slit. The pipeline extracts 1D spectra for each object using both a boxcar and an optimal extraction algorithm \citep{Horne_1986}; we chose to use the optimally extracted spectra in all that follows. 

\subsection{VLT}
\label{s_VLTReduction}
The FORS2 detector consists of two 2k by 4k E2V CCDs. Each chip was processed separately. The bias subtraction, flat-fielding and wavelength calibration of the FORS2 data were done in a standard manner. The bias was estimated by fitting low order polynomials to the overscan regions. The flat-fields were created from the lamp flats and the wavelength calibration was done with arc frames. Several bright OH lines were used to check the wavelength calibration. There were no systematic offsets and the RMS scatter in the offset was 0.4 \AA{}, which corresponds to approximately 1/10th of a pixel.

The removal of the sky was tried in two ways. First, the sky was removed by subtracting low order polynomials along the spatial direction of the spectrum. If the redshift could not be clearly determined, usually because the signal from the object was dominated by detector fringing, the sky was removed by subtracting a two dimensional sky frame that was created from the data itself. The second method removes the fringes at the expense of slightly reducing the signal-to-noise ratio of the extracted spectra.

The process of creating a two dimensional sky frame is complicated by the fact that the spectrum of the night sky changes with time. Taking a simple median of the two dimensional spectra and then subtracting the result is generally unsatisfactory. Instead, we combine the two dimensional spectra and perform the sky subtraction on a column-by column basis, allowing for the fact that the sky lines on the two dimensional spectra are slightly curved and that their intensity varies with time. This method also allows one to find and exclude cosmic rays at the same time. The method is described in greater detail in the FORS Data Reduction Cookbook\footnotemark.
\footnotetext{Available from http://www.eso.org/instruments/fors/.} 

%=============================================================================================
\section{Analysis}
\label{s_SpectroAnalysis}
Redshifts were measured from the DEIMOS and FORS2 spectra using the Fourier cross-correlation technique of \citet{Tonry_1979}. This was implemented using the task \texttt{xcsao} in version 2.4.9 of the \texttt{rvsao} radial velocity package \citep{Kurtz_1998} for the \texttt{IRAF}\footnotemark environment. 
\footnotetext{IRAF is distributed by the National Optical Astronomy Observatories, which are operated by the Association of Universities for Research in Astronomy, Inc., under cooperative agreement with the National Science Foundation.}

We correlated the spectra with the SDSS spectral templates\footnotemark, which have a typical wavelength coverage of $\sim$3800-9200 \AA{}, and the emission line template supplied in the \texttt{rvsao} package. We matched against a subset of stellar templates (covering the whole range of spectral types), the full set of available galaxy templates, and a QSO template. We supplemented the SDSS templates with an additional Luminous Red Galaxy (LRG) template constructed by \citet{Eisenstein_2003}, which provides coverage over 2600-4300 \AA{} in the rest frame. In the case of the DEIMOS data, we found it necessary to produce a customized emission line template featuring a split 3727 \AA{} [O \textsc{ii}] line, using the \texttt{rvsao} task \texttt{linespec}, in order to significantly reduce the number of template misidentifications of genuine high-redshift [O \textsc{ii}] emission with low-redshift H$\alpha$ emission. To remove bright sky lines, which are generally a significant feature at the red-end of the object spectra, we use the ability of \texttt{xcsao} to mask out and replace user-defined regions with a linear interpolation.
\footnotetext{http://www.sdss.org/dr5/algorithms/spectemplates/index.html}

\begin{figure}
\epsscale{0.9}
\plotone{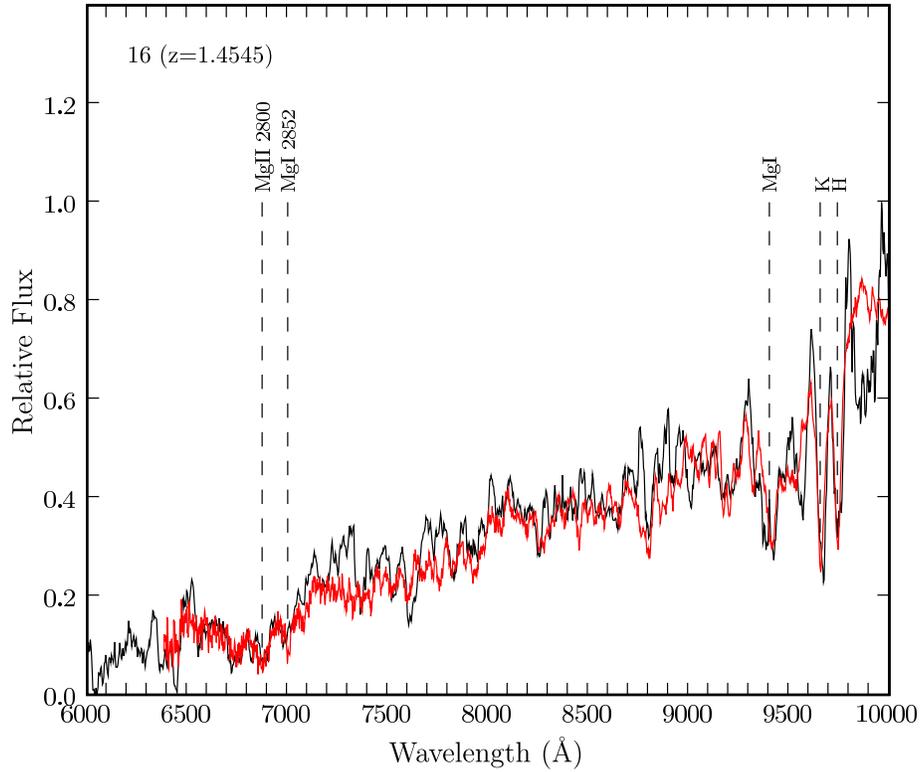}
\caption{Example 1D FORS2 spectrum (smoothed by a 10 pixel boxcar) of a galaxy identified as a member of J2215.9$-$1738. The Luminous Red Galaxy (LRG) spectral template of \citet{Eisenstein_2003}, which was cross-correlated with the object spectrum in order to obtain the redshift measurement, is overlaid in red.}
\label{f_exampleSpectrum}
\end{figure}

Typically, each DEIMOS mask contains $\sim100$ slitlets, so we found it desirable to automate the redshift measurements as much as possible. The results of \texttt{xcsao} are dependent upon the value of the input initial trial redshift, and so we vary this initial redshift over the range $0.0 < z < 1.8$ in steps of 0.2 and record the measured redshift and associated goodness-of-fit value $R$. At the end of this process, we visually inspected every object spectrum, using \texttt{xcsao} to mark the appropriate spectral features at the best fitting cross-correlation redshift. In this way we were able to reject some cases of a spurious best fitting redshift in favor of a correctly measured redshift recorded with a lower $R$ value. Fig.~\ref{f_exampleSpectrum} shows an example VLT object spectrum with the redshifted \citet{Eisenstein_2003} LRG spectral template overlaid. In some cases the cross-correlation technique did not yield the correct redshift for the object spectrum. This occurred most frequently for FORS2 spectra of objects with 3727 \AA{} [O \textsc{ii}] emission. In such cases we measured the object redshift from the centroids of visually identified lines. Uncertainties on redshifts measured with the cross-correlation technique are typically $\Delta z \sim 10^{-4}$, as estimated by \texttt{xcsao}. In the case of redshifts measured from the centroids of spectral lines, the size of the error is estimated visually: in the case of FORS2 spectra of galaxies identified by [O \textsc{ii}] emission, the accuracy with which redshifts can be measured is limited to $\Delta z \sim 10^{-3}$ by the spectral dispersion per pixel. The technique used to obtain the redshift measurement for each galaxy is noted in Table~\ref{t_J22159members}.

All measured redshifts were assigned a quality flag $Q$ according to the following system: $Q=3$ (completely unambiguous, at least two positively identified spectral features); $Q=2$ (high confidence that the redshift is correct, one clearly detected feature); $Q=1$ (significant doubt that the redshift is correctly identified, one or more weakly detected features). Note that the spectral resolution of the DEIMOS data is sufficient to resolve the two components of the 3727 \AA{} [O \textsc{ii}] line -- in cases where this is clearly visible, [O \textsc{ii}] is counted as two spectral features (i.e., such spectra are assigned $Q=3$). We consider only galaxies with $Q \geq 2$ redshifts in the subsequent discussion.

%=============================================================================================
\section{Results}
\label{s_SpectroResults}
\label{s_x2215VelocityDispersion}
\begin{figure}
\epsscale{0.9}
\plotone{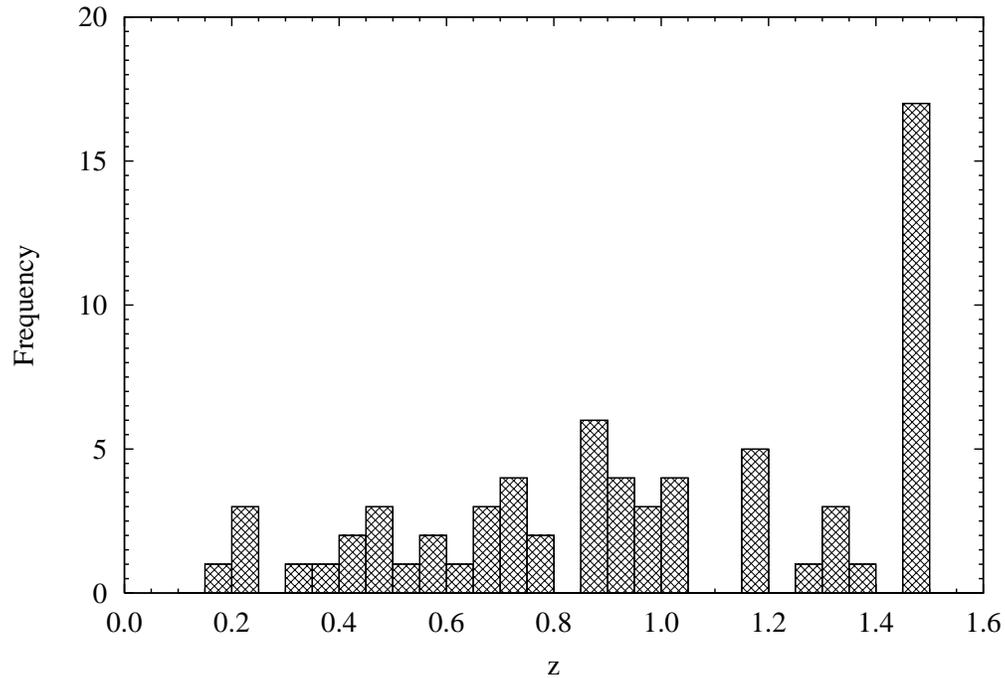}
\caption{Distribution of $Q \geq 2$ galaxy redshifts (in bins of width 0.05 in $z$) within a 2$\arcmin$ radius of the X-ray position of J2215.9$-$1738. The cluster is clearly identified by the peak at $z=1.45$.}
\label{f_LoS_zHistogram}
\end{figure}

Fig.~\ref{f_LoS_zHistogram} shows the redshift distribution of galaxies with $Q \geq 2$ redshifts located within a $2\arcmin$ radius of the cluster X-ray position. The cluster is clearly identified as the peak in the redshift distribution at $z = 1.45$. In addition to the six members reported in \citet{Stanford_2006}, the new spectroscopic observations yielded a further 15 galaxies with $Q \geq 2$ redshifts in the range $1.44 < z < 1.48$ within 2 Mpc of the cluster X-ray position. Table~\ref{t_J22159members} presents a list of the cluster members. 

We calculated an initial estimate of the cluster redshift using a biweight location estimator \citep{BeersBiweight_1990}, applied to all 21 galaxies with $Q \geq 2$ listed in Table~\ref{t_J22159members}, obtaining a value of $z=1.459 \pm 0.002$, where the uncertainty is estimated using a bootstrap resampling technique. Similarly applying a biweight scale estimator \citep{BeersBiweight_1990} to obtain an initial estimate of the cluster line of sight velocity dispersion, we find $\sigma_v=840 \pm 150$ km s$^{-1}$. However, no selection criteria in either velocity or radial distance from the cluster center were used to derive these initial estimates. More robust estimates of these quantities can be obtained by restricting the membership to those objects found within the virial radius $R_{\rm v}$, as by definition, galaxies within $R_{\rm v}$ should be gravitationally bound within the cluster potential well. $R_{\rm v}$ can be estimated using our knowledge of the X-ray properties of J2215.9$-$1738 and some assumptions. The self-similar evolution of $R_{\rm v}$, defined with respect to the critical density \citep[see][]{VoitScaling_2005}, depends only on the cluster X-ray temperature and the value of the Hubble parameter at the cluster redshift $z$ , i.e.,

\begin{equation}
\label{e_calcRV}
R_{\rm v} = 3.80 \beta_T^{1/2} E(z)^{-1} \left( kT/10 {\ \rm keV} \right)^{1/2} h_{50}^{-1}, {\rm \quad where} 
\end{equation}
\\
\begin{equation}
\label{e_EZ}
E(z)=[\Omega_m(1+z)^{3}+(1-\Omega_m-\Omega_\Lambda)(1+z)^2+\Omega_\Lambda]^{1/2}.
\end{equation}
\\
$\beta_T$ in equation~\ref{e_calcRV} is the normalization of the virial relation $GM_{\rm v}/2R_{\rm v}=\beta_T kT$, for which we adopt the value 1.05 \citep{Evrard_1996}. $E(z)$ describes the evolution with redshift of the Hubble parameter.

Using equation~\ref{e_calcRV}, the virial radius was estimated to be 1.05 Mpc at $z=1.459$, for the measured temperature of $kT=7.4$ keV \citep{Stanford_2006}. This distance is equivalent to an angular distance on the sky of $\simeq 2.1 \arcmin$. This estimate of $R_{\rm v}$ was then used to select cluster members and refine the measurement of the cluster redshift: applying the biweight location estimator to the 17 galaxies located within $R_{\rm v}$ as listed in Table~\ref{t_J22159members}, we find that the cluster redshift is $z=1.457 \pm 0.002$. We expect this to be a more robust estimate of the cluster redshift than the initial estimate, as galaxies within $R_{\rm v}$ should be gravitationally bound within the cluster. 

We determined the cluster line of sight velocity dispersion in an iterative fashion. Initially, galaxies within $R_{\rm v}$ were selected with peculiar velocities within $\pm 2000$ km s$^{-1}$ of the cluster velocity corresponding to $z=1.457$, and on subsequent iterations a sigma clipping algorithm was used to discard galaxies with velocities outside of $\pm 3 \times \sigma_v$ (although in practice, the conservative clipping applied meant that no galaxies were in fact removed from the sample in this way). We obtain a result of $\sigma_v = 580 \pm 140$ km s$^{-1}$ in the rest-frame from 17 members, where the uncertainty is estimated using a bootstrap resampling technique. Note we have subtracted a contribution of $\approx 90$ km s$^{-1}$ in order to take into account broadening of the velocity distribution by uncertainties in the redshift measurements, following \citet*{Danese_1980}. If, alternatively, we select galaxies within a fixed radius of 2 Mpc instead of using the virial radius, we obtain $\sigma_v=620 \pm 120$ km s$^{-1}$ from 19 galaxies by the same method. The value of $\sigma_v$ we obtain for J2215.9$-$1738 is of similar size to the only other $z > 1.3$ cluster so far discovered, XMMU J$2235.3-2557$ at $z=1.39$, which has $\sigma_v=762 \pm 265$ km s$^{-1}$ \citep{Mullis_2005}. 

\begin{figure}
\epsscale{0.9}
\plotone{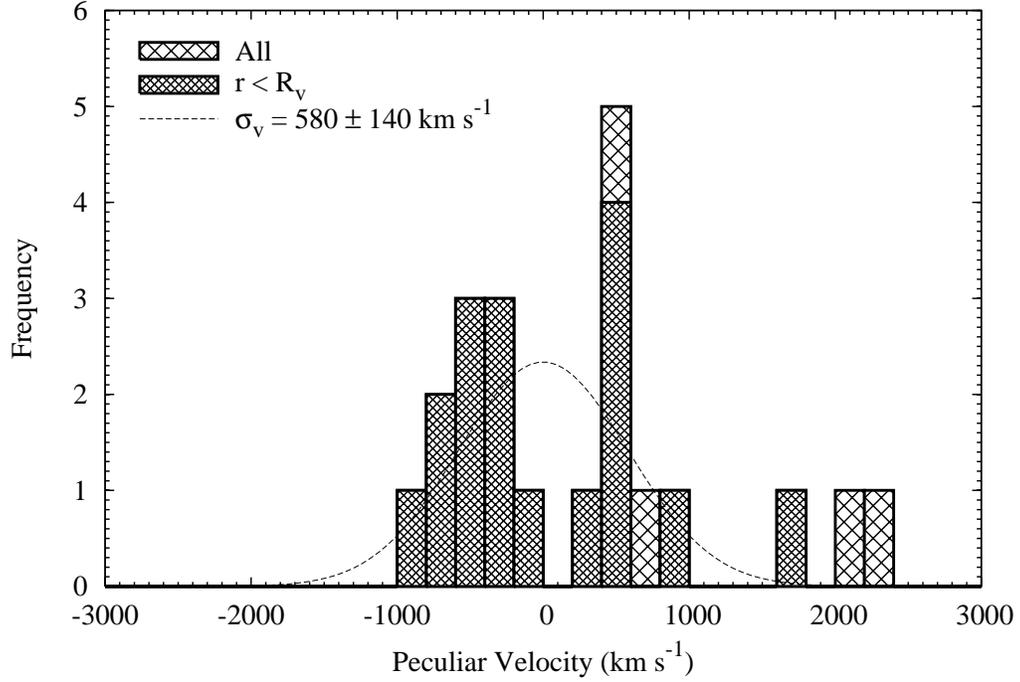}
\caption{Velocity histogram (in 200 km s$^{-1}$ bins), centered on the velocity corresponding to the adopted cluster redshift of $z=1.457$ (see the text in \textsection~\ref{s_SpectroResults}). Only galaxies with $Q \geq 2$ redshifts are plotted. The shading indicates as appropriate: galaxies located within $\pm 2000$ km s$^{-1}$ of the cluster redshift and inside the virial radius ($r < R_{\rm v}$), as determined from the X-ray temperature (equation~\ref{e_calcRV}); and all 21 galaxies with $Q \geq 2$ redshifts listed in Table~\ref{t_J22159members}. The dashed line is a Gaussian distribution with standard deviation equal to the velocity dispersion calculated using galaxies within $R_{\rm v}$, normalized to the total number of objects within this sample.}
\label{f_J22159pecvelhist}
\end{figure}

\begin{figure}
\epsscale{0.9}
\plotone{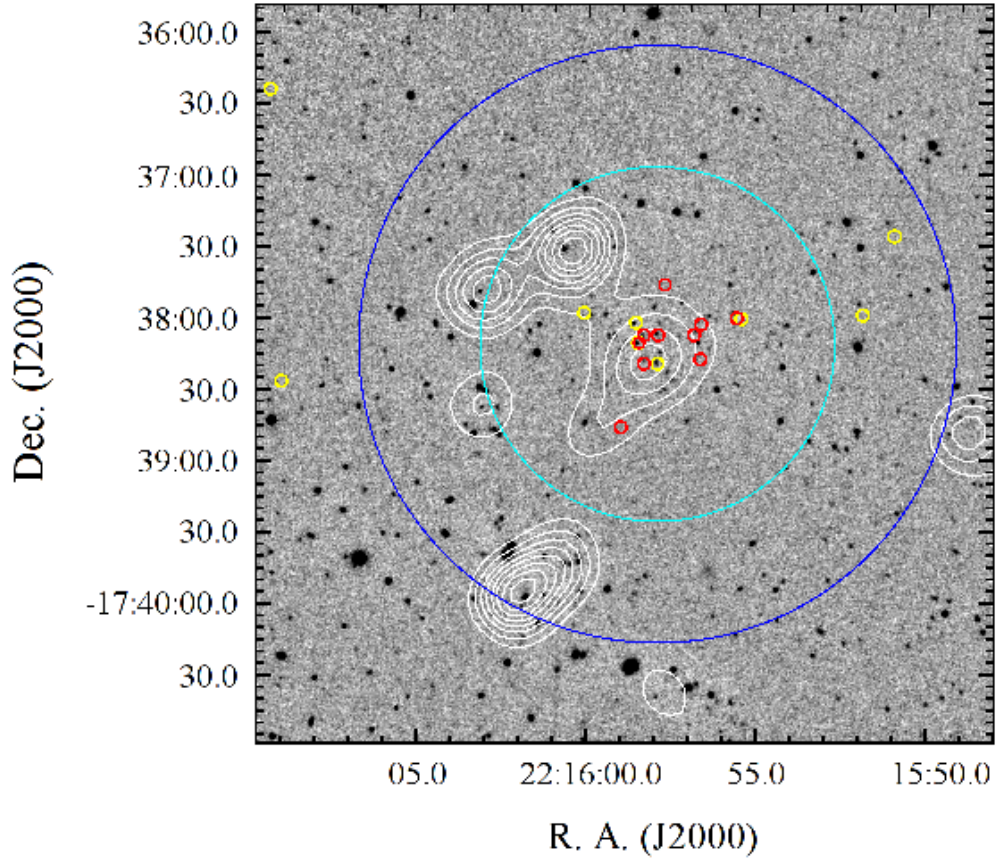}
\caption{$5.2 \arcmin \times 5.2 \arcmin$ $K_s$-band image of J2215.9$-$1738, with X-ray contours overlaid in white. Objects spectroscopically identified as cluster members with $Q \geq 2$ within $3 \times$ the velocity dispersion of the cluster and within a projected 2 Mpc radius are highlighted (see text). Members with $z<1.457$ are highlighted in red; members with $z>1.457$ are highlighted in yellow. There is no clear spatial separation between the two sets of highlighted objects, indicating that any substructure must be along the line of sight. Note that some members are not detected in the $K_s$-band image. The dark blue circle marks the cluster virial radius of 1.05 Mpc, as calculated using equation~\ref{e_calcRV}; the light blue circle marks the radius $R_{200}=0.63$ Mpc, calculated from the cluster velocity dispersion using equation~\ref{e_R200}.}
\label{f_J22159members}
\end{figure}

In Fig.~\ref{f_J22159pecvelhist} we plot the velocity distribution of all the objects with $Q \geq 2$ redshifts listed in Table~\ref{t_J22159members}, centered on the velocity corresponding to $z=1.457$. The 17 members identified to lie within the virial radius of the cluster are highlighted. At first glance, the velocity distribution of the galaxies located within $R_{\rm v}$ appears bimodal. However, as shown by Fig.~\ref{f_J22159members}, there is no clear separation of the two velocity subclumps in the plane of the sky: therefore if substructure is present, it must be aligned close to the line of sight. The Shapiro-Wilks test indicates that the peculiar velocity distribution is marginally consistent with being drawn from a Gaussian distribution, at the 10 per cent level. We also performed the dip test of unimodality \citep{Hartigan2_1985} on the velocity distribution, obtaining a value for the dip statistic of 0.0962. There is a $< 20$ per cent probability of obtaining a dip value larger than this value for a sample size of 17 when drawing from a uniform distribution. As a further test, we used Monte Carlo simulations to determine the probability of obtaining a value for the dip statistic as large as measured by drawing 17 values at random from a single Gaussian distribution. From 10000 realizations, we determine that there is a $6$ per cent probability of obtaining a dip statistic larger than 0.0962, when drawing from a single Gaussian distribution with standard deviation equal to the cluster velocity dispersion. 

We conclude that on the basis of the present data there is mild evidence that the cluster velocity distribution deviates from that expected from a single Gaussian. Though it is very unlikely that the possible bimodal velocity distribution shown in Fig.~\ref{f_J22159pecvelhist} is the result of a selection effect given the simple colour--magnitude criteria used to select target galaxies (\textsection~\ref{s_SpectroObservations}, above), it may arise due to incompleteness, given the small number of objects in the sample. More data are clearly needed to confirm if significant substructure is present.

%=============================================================================================
\section{Discussion}
\label{s_SpectroDiscuss}
J2215.9$-$1738 is the most distant cluster currently known for which measurements of the X-ray luminosity, temperature, and line of sight galaxy velocity dispersion are available. We now ask the question, are the observed cluster properties consistent with those expected from studies of X-ray scaling relations at low redshift?

%---------------------------------------------------------------------------------------------
\subsection{The $L_{\rm X}-T$ relation at $z \sim 1.5$}
\label{s_Discuss_LT}
It has been known for some time that the slope of the low redshift $L_{\rm X}-T$ relation is inconsistent with that expected from self-similar models (i.e., where gravitational processes are solely responsible for setting the observed X-ray properties of clusters), which predict $L_{\rm X} \propto T^2$ \citep{Kaiser_1986}. Many studies of the $L_{\rm X}-T$ relation have shown consistently that the observed slope of the relation is closer to $L_{\rm X} \propto T^3$ \citep[e.g.,][]{David_1993, Markevitch_1998, ArnaudEvrard_1999}, indicating that some form of non-gravitational heating has taken place within clusters. However, several studies using cluster samples up to $z\sim 1$ have found that although the slope of the $L_{\rm X}-T$ relation departs from the expected self-similar value, the \textit{evolution} of the relation is nevertheless consistent with self-similarity, although the scatter in such measurements is large \citep[e.g.][]{Maughan_2006}. The self-similar evolution of the observed local bolometric $L_{\rm X}-T$ relation can be expressed as,

\begin{equation}
\label{e_selfsimilarLXT}
E(z)^{-1} L_{\rm X}=6.35 \left( {\displaystyle\frac{kT}{6 {\rm \ keV}}} \right) ^{2.64},
\end{equation}
\\
where $E(z)$ is given by equation~\ref{e_EZ}. The slope of 2.64 adopted in equation~\ref{e_selfsimilarLXT} is the slope of the local $L_{\rm X}-T$ relation as measured by \citet{Markevitch_1998}. The units of $L_{\rm X}$ in equation~\ref{e_selfsimilarLXT} are $10^{44}$ ergs s$^{-1}$.

Using equation~\ref{e_selfsimilarLXT} to predict the expected X-ray luminosity of J2215.9$-$1738, assuming the measured cluster temperature of $kT = 7.4^{+1.6}_{-1.1}$ keV, we obtain $L_{\rm X} \sim (1.6-4.2) \times 10^{45}$ ergs s$^{-1}$, neglecting the scatter in the relation. This is significantly larger than the measured luminosity of $4.39^{+0.46}_{-0.37} \times 10^{44}$ ergs s$^{-1}$. The discrepancy is naturally reduced if we adopt $kT = 6.5^{+1.6}_{-1.1}$ keV, the measured temperature if an undetected central point source is present \citep{Stanford_2006}, but the measured luminosity is still significantly lower than that which would be expected from self-similar evolution of the local $L_{\rm X}-T$ relation.

A possible solution to this discrepancy is that the evolution of the $L_{\rm X}-T$ relation is more accurately described with the inclusion of the effects of non-gravitational heating and radiative cooling \citep[see, e.g., the review by][]{Voit_2005}. Radiative cooling has the effect of raising the temperature of the intracluster medium (ICM), because selectively removing the low-temperature gas leads to a higher average temperature for the remaining material in the gas phase \citep{Voit_2002}. However, cooling is a runaway process that must be regulated by some form of heating, otherwise the cooling gas should condense to produce significant amounts of star formation in cluster cores -- which is not what is observed \citep[e.g.,][]{Balogh_2001}. Heating of the ICM by supernovae, star formation and/or Active Galactic Nuclei (AGN) is thought to provide the feedback mechanism required to prevent overcooling, and has been postulated for some time as the solution to the well-known `cooling flow' problem in clusters \citep[see, e.g., the review by][]{Fabian_1994}. Hydrodynamical simulations including the effects of radiative cooling, star formation, and supernovae feedback are able to reproduce the observed $L_{\rm X}-T$ relation quite well, finding a slope in the range 2.5-3 \citep[e.g.,][]{Kay_2007, Borgani_2004}. However, the redshift evolution of the normalization of the relation when such effects are taken into account is quite different to the self-similar case, in which clusters are expected to become more luminous for a given temperature as redshift increases; in particular, simulations that include the effects of feedback show a mildly negative evolution as redshift increases \citep{Muanwong_2006}.

\begin{figure}
\epsscale{0.9}
\plotone{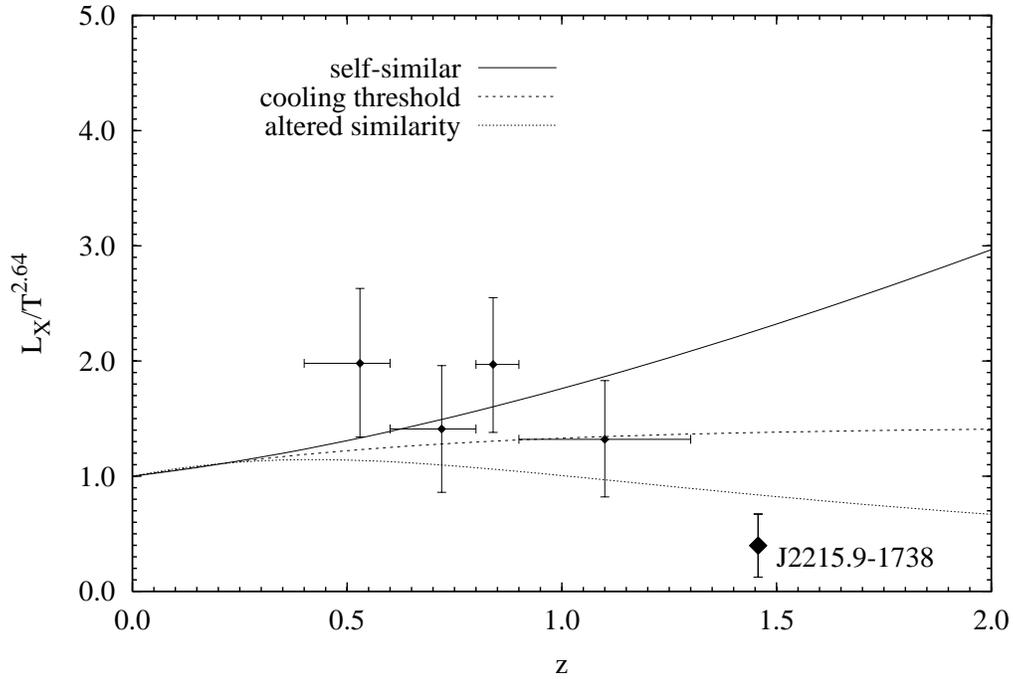}
\caption{Comparison of $L_{\rm X}/T^{2.64}$ for J2215.9$-$1738 with predicted evolution of the normalization of the $L_{\rm X}-T$ relation for the cases of self-similarity ($E(z)$), cooling threshold ($t_0 / \left[ E(z) t(z) \right] $), and altered similarity ($t_0^2 / \left[ E(z)^3 t(z)^2 \right] $). J2215.9$-$1738 is clearly more consistent with these latter two models, which are representative of the form of evolution expected when the effect of feedback on the intracluster medium is taken into account. The points are the data of \citet{Maughan_2006}; vertical error bars are equal to the weighted standard deviation at each redshift, horizontal error bars indicate the width of each redshift bin.}
\label{f_LTnormalisation_evo}
\end{figure}

\citet{VoitScaling_2005} provides analytic predictions for the evolution of the $L_{\rm X}-T$ relation normalization derived from a semi-analytic framework that includes the effects of both non-gravitational heating and radiative cooling. The modified entropy models of \citet{Voit_2002} show that introducing a cooling threshold, where gas with entropy less than $K_c = T^{2/3} t(z)^{2/3}$ radiates away its thermal energy within a timescale equal to the age of the universe $t(z)$, can steepen the expected $L_{\rm X}-T$ relation to $L_{\rm X} \propto T^{2.5}$, closer to the observed value at low redshift. Furthermore, the predicted evolution of the $L_{\rm X}-T$ relation in this scenario is quite different to the self-similar case:

\begin{equation}
\label{e_coolingThreshold}
L_{\rm X} \propto kT^{2.5} E(z)^{-1} t(z)^{-1}.
\end{equation}
\\
The additional inclusion of the effect of smooth accretion of gas by clusters modifies the expected form of the $L_{\rm X}-T$ relation to $L_{\rm X} \propto T^3$, and leads to the so called `altered similarity' evolution of the relation \citep{VoitPonman_2003, Voit_2003}: 

\begin{equation}
\label{e_alteredSimilarity}
L_{\rm X} \propto kT^{3} E(z)^{-3} t(z)^{-2}.
\end{equation}
\\
Fig.~\ref{f_LTnormalisation_evo}, based on Fig. 14 of \citet{Maughan_2006}, shows the position of J2215.9$-$1738 (assuming the \citealt{Markevitch_1998} $L_{\rm X}-T$ relation) relative to the self-similar (equation~\ref{e_selfsimilarLXT}), cooling threshold (equation~\ref{e_coolingThreshold}), and altered similarity (equation~\ref{e_alteredSimilarity}) predictions for the evolution of the $L_{\rm X}-T$ relation normalization. The data points in Fig.~\ref{f_LTnormalisation_evo} are taken from \citet{Maughan_2006}, and represent the weighted mean values (assuming the \citealt{Markevitch_1998} relation) in each redshift bin for their sample of 33 clusters. The error bars are the weighted standard deviation, and illustrate the large amount of scatter in $L_{\rm X}/T^{2.64}$. Nevertheless, it can be seen that the position of J2215.9$-$1738 is more consistent with the cooling threshold and altered similarity evolution predictions than with self-similarity. It is clear from Fig.~\ref{f_LTnormalisation_evo} that observations of much larger samples of clusters at high redshift are required if the evolution of the $L_{\rm X}-T$ relation is to be adequately constrained.
 
A possible alternative explanation to the relative faintness of the cluster X-ray emission in comparison to its temperature is that the cluster has undergone a merger event in its recent past. As shown in \textsection~\ref{s_SpectroResults}, although the velocity distribution of J2215.9$-$1738 is marginally consistent with a single Gaussian distribution (at the $\sim10$ per cent level), there is marginal evidence (at the $<2 \sigma$ level) from the \citet{Hartigan2_1985} dip test in favor of a bimodal distribution. Clearly, there is a strong possibility that further spectroscopic observations may reveal the presence of significant substructure within the cluster, and this would not be unexpected in the hierarchical structure formation scenario, as the frequency of cluster mergers is expected to increase with redshift. The numerical simulations of \citet{Poole_2007} \citep[see also][]{RickerSarazin_2001} show that although both $L_{\rm X}$ and $T$ are boosted significantly for approximately a sound crossing time during the first pericentric passage of two initially relaxed clusters, in the case of an off-axis merger the resulting remnant can have luminosity up to $50$ per cent lower than would be expected from the observed cluster mass--scaling relations $\sim$a few Gyr after the encounter, with $T$ boosted by $\sim 10$ per cent. Assuming that this is the case, the discrepancy between the observed luminosity and that implied by equation~\ref{e_selfsimilarLXT} (assuming $kT=7.4$ keV) falls to $\sim 2\sigma$.

%---------------------------------------------------------------------------------------------
\subsection{The $L_{\rm X}-\sigma_{\lowercase{v}}$ relation at $z \sim 1.5$}
\label{s_Discuss_Lsigma}
Under the assumptions that clusters are virialized, with isothermal gas and galaxy distributions, and that the gas mass bound to clusters is proportional to the virial mass, then the expected relationship between X-ray luminosity and galaxy velocity dispersion is $L_{\rm X} \propto \sigma_v^4$ \citep{QuintanaMelnick_1982}. This relation has been measured for clusters in the local universe and generally found to be consistent with the expected result \citep[e.g.,][]{Ortiz-Gil_2004, Hilton_2005}, albeit with large scatter. In the case where the local relation does not evolve, using the $L_{\rm X}-\sigma_v$ relation derived by \citet{Ortiz-Gil_2004}, i.e. $L_{\rm X} \propto \sigma_v^{4.1 \pm 0.3}$, we expect the bolometric X-ray luminosity of J2215.9$-$1738 to be $L_{\rm X} \sim (1.1^{+1.6}_{-0.7}) \times 10^{44}$ ergs s$^{-1}$.

We can compare this result with that expected if the evolution of the relation is self-similar by modifying the \citet{Ortiz-Gil_2004} relation to the form,

\begin{equation}
\label{e_selfsimilarLXSBI}
E(z)^{-1} L_{\rm X} =  10^{32.72} \times \sigma_v^{4.1},
\end{equation}
\\
where $L_{\rm X}$ is in units of ergs s$^{-1}$, and $\sigma_v$ is in units of km s$^{-1}$.

Accounting for self-similar evolution using equation~\ref{e_selfsimilarLXSBI}, we find that for $\sigma_v=580 \pm 140$ km s$^{-1}$, we expect $L_{\rm X} \sim (2.6^{+3.6}_{-1.8}) \times 10^{44}$ ergs s$^{-1}$. Due to the large uncertainty in $\sigma_v$, both the no-evolution and self-similar evolution cases are consistent with the measured X-ray luminosity of J2215.9$-$1738, and we are therefore unable to discriminate between them.

%---------------------------------------------------------------------------------------------
\subsection{The $\sigma_{\lowercase{v}}-T$ relation at $z \sim 1.5$}
The assumption of isothermality for both the gas and galaxy velocity distributions leads to the parametrization $\beta=\sigma_v^2 \mu m_p / kT$, where $\mu$ represents the mean molecular weight and $m_p$ is the proton mass. By adopting $\beta=1$, we can test whether the measured velocity dispersion and temperature are consistent with equipartition of energy between the gas and galaxies. This is expected in the case of self-similar evolution, where the gas temperature and galaxy velocity dispersion are assumed to both perfectly trace the depth of the gravitational potential well. \citet*{WuXueFang_1999} measured the $\sigma_v-T$ relation from a heterogeneous sample of 92 clusters drawn from the literature, finding it to be consistent with $\beta \approx 0.9-1.0$.

We find that the predicted X-ray temperature for $\sigma_v=580 \pm 140$ km s$^{-1}$ is $kT_{\rm pred}=2.7^{+1.3}_{-1.0}$ keV (assuming $\mu=0.58$, \citealt*{EdgeStewart_1992}). The difference between the measured X-ray temperature of $kT=7.4^{+1.6}_{-1.1} $ keV and $kT_{\rm pred}$ is $\Delta kT=5.4 \pm 1.7$ keV, a $>3 \sigma$ discrepancy. If we assume that the cluster contains an unresolved central X-ray point source, then the measured X-ray temperature is $kT=6.5^{+1.6}_{-1.1}$ keV (\citealt{Stanford_2006}), and the discrepancy with $kT_{\rm pred}$ falls to $2-3 \sigma$. J2215.9$-$1738 therefore appears to deviate significantly from the $\sigma_v-T$ relation. Another way of stating this is to calculate $\beta$ directly from the measured velocity dispersion and temperature. We find that for $kT=7.4$ keV and $\sigma_v=580$ km s$^{-1}$, $\beta = 0.28 \pm 0.14$. In the case of $kT=6.5$ keV, we obtain a consistent result, $\beta = 0.31 \pm 0.16$. This indicates that the gaseous component of the cluster has $\sim 2-3$ times the kinetic energy per unit mass than the galaxies. 

The value of $\beta$ we obtain for this cluster is therefore incompatible with the expectation from self-similar evolution, for which, by definition, $\beta =1$ at all redshifts. This suggests that the virial radius used to select cluster members (equation~\ref{e_calcRV}), has been overestimated. We can test the effect of this upon our results by calculating $\sigma_v$ within a different characteristic radius defined independently of the X-ray temperature,

\begin{equation}
R_{200}=\frac{\sqrt{3} \sigma_v}{10H(z)},
\label{e_R200}
\end{equation} 
\\
where $R_{200}$ is the radius that encloses a mean density 200 times the critical density at a given redshift \citep[roughly equivalent to the virial radius,][]{CarlbergR200_1997}, and $H(z)$ is the Hubble parameter at redshift $z$. 

Using equation~\ref{e_R200} with the initial estimate of $\sigma_v=840 \pm 150$ km s$^{-1}$, obtained with no restriction on the selection of cluster members in radial distance (i.e. using all 21 galaxies with $Q \geq 2$ redshifts listed in Table~\ref{t_J22159members}, see \textsection~\ref{s_SpectroResults} above), we find $R_{200}=0.91 \pm 0.15$ Mpc. Calculating the cluster redshift and velocity dispersion using members selected within this radius in a similar fashion to \textsection~\ref{s_SpectroResults} above, we obtain $z=1.456 \pm 0.002$ and $\sigma_v=580 \pm 190$ km s$^{-1}$ respectively from 16 members. This is consistent with the results found in \textsection~\ref{s_SpectroResults} when selecting members within $R_{\rm v}$ as estimated using the cluster X-ray temperature, though the larger uncertainty in $\sigma_v$ in this case reduces the discrepancy between the temperature implied by the velocity dispersion and the measured X-ray temperature to the $2-3 \sigma$ level. Further iterations in estimating $R_{200}$ and $\sigma_v$ in this manner naturally return consistent results, converging to $R_{200}=0.63 \pm 0.15$ Mpc, $\sigma_v=570 \pm 190$ km s$^{-1}$ from 15 members.

We now consider ways in which $T$ could be increased above its expected value, or equivalently, how $\sigma_v$ could have been reduced. As stated in \textsection~\ref{s_Discuss_LT}, one way to boost $T$ significantly is through cluster mergers \citep{RickerSarazin_2001, Poole_2007}. This is a possibility, as there is marginal evidence that the observed galaxy velocity distribution is inconsistent with a single Gaussian (\textsection~\ref{s_SpectroResults}). Another way in which our result could be reconciled with the self-similar expectation is if the cluster exhibits a significant velocity distribution anisotropy. However, the required anisotropy would be large: from the $\sigma_v-T$ relation measured by \citep{WuXueFang_1999}, we require $\sigma_v \sim 1100$ km s$^{-1}$ to reproduce the measured X-ray temperature of $7.4$ keV, which is significantly larger than the measured velocity dispersion.

\begin{figure}
\epsscale{0.9}
\plotone{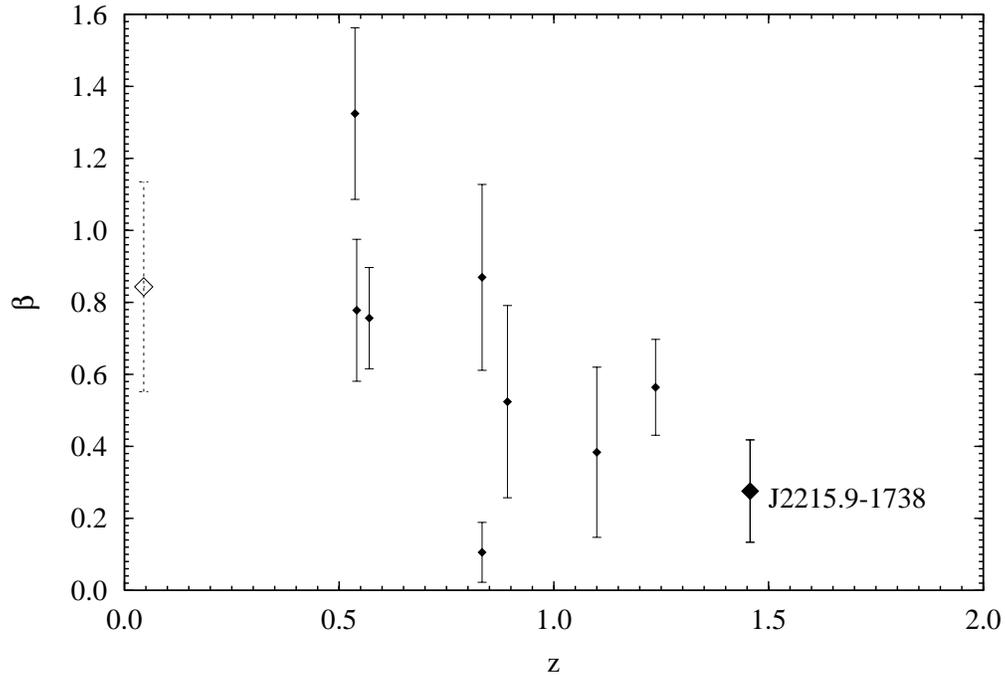}
\caption{Comparison of J2215.9$-$1738 with calculated values of $\beta$ versus redshift for a heterogeneous sample of clusters at $z>0.5$ drawn from the literature, for which the uncertainty in $\beta <0.3$. At high redshift, there appears to be a tendency for $\beta$ to be lower than expected for self-similar evolution of the intracluster medium. The open point is the mean $\beta$ (with error bars equal to the standard deviation) for 59 clusters at $z<0.1$ drawn from the sample of \citet{WuXueFang_1999}.}
\label{f_sigmaTnormalisation_evo}
\end{figure}

We note that J2215.9$-$1738 is not the only high-redshift cluster to possess a low $\beta$ value. \citet*{LubinMulchaeyPostman_2004} extended the $\sigma_v-T$ relation significantly in redshift using data drawn from the literature for 11 clusters at $z>0.5$. They noted that although the $z>0.5$ clusters are consistent within the errors of the local $\sigma_v-T$ relation, their temperatures are higher by a factor of $\sim 1.4$ on average for a given velocity dispersion. Most recently, \citet{Demarco_2007} carried out a study of the dynamical structure of RDCS J1252.9$-$2927 at $z=1.237$, and found that the temperature implied by the galaxy velocity dispersion (estimated using the $\sigma_v-T$ relation) is lower than the measured X-ray temperature at $>3 \sigma$ significance. In Fig.~\ref{f_sigmaTnormalisation_evo}, we plot calculated values of $\beta$ versus redshift for objects drawn from the $z>0.5$ cluster sample of \citet{LubinMulchaeyPostman_2004}, supplemented by other high-z clusters for which measurements of velocity dispersion and temperature have appeared in the literature since this work. Only objects where the uncertainty in $\beta$ is $<0.3$ are plotted (errors are combined in quadrature), and the sample used is listed in Table~\ref{t_SigmaTLiterature}. For comparison, we also plot the mean $\beta$ (with error bars equal to the standard deviation) for 59 clusters at $z<0.1$ drawn from the  \citet{WuXueFang_1999} sample, where again, only clusters for which the uncertainty in $\beta$ is $<0.3$ have been included.

Fig.~\ref{f_sigmaTnormalisation_evo} shows a trend of $\beta$ decreasing towards higher redshift, although there are few clusters with velocity dispersions and temperatures measured with reasonable precision at high redshifts. As at least two of the clusters in the literature sample (RDCS J1252.9$-$2927 and RX J0152.7$-$1357) are confirmed multi-component systems, the trend in Fig.~\ref{f_sigmaTnormalisation_evo} may represent the increasing frequency of cluster mergers expected at high redshifts in the hierarchical structure formation scenario. Alternatively, as it is well known that non-gravitational processes must affect the evolution of the ICM in order to explain the observed $L_{\rm X}-T$ relation at low-redshift, we may suppose that these same processes could result in the evolution of the $\sigma_v-T$ relation with redshift. It is easy to imagine that heating by supernovae or AGN, for example, could modify the X-ray temperature above the self-similar expectation, but it is difficult to think of a process which could act to reduce the galaxy velocity dispersion significantly. A comparison with the properties of clusters in the \textit{Millennium Gas} simulations (Pearce et al., in preparation), which have sufficient volume to resolve $kT > 7$ keV clusters at $z > 1$, should provide insight into which process is responsible for the observed properties of J2215.9$-$1738.

%=============================================================================================
\section{Conclusions}
\label{s_SpectroConclusions}
We have increased the number of known members of the most distant galaxy cluster known, XMMXCS J2215.9$-$1738 at $z=1.457$, to 17 objects located within the virial radius as estimated using the cluster X-ray temperature, and have measured its line-of-sight velocity dispersion to be $\sigma_v = 580 \pm 140$ km s$^{-1}$. For the measured X-ray temperature of $7.4^{+1.6}_{-1.1}$ keV \citep{Stanford_2006}, this is inconsistent with the hypothesis of equipartition of energy between the gas and the galaxies ($\beta=\sigma_v^2 \mu m_p / kT = 1$) at the $\approx 3 \sigma$ level, and indicates that the intracluster medium contains $2-3$ times the kinetic energy of the galaxies. The cluster X-ray emission is significantly fainter than expected from self-similar evolution of the local observed $L_{\rm X}-T$ relation. 

The cluster properties could be the result of the cluster undergoing a merger within the last few Gyr, although we find only mild evidence from the present data that the velocity distribution is inconsistent with being drawn from a single Gaussian. An alternative possibility is that the effect of heating of the intracluster medium by supernovae and/or AGN is responsible. Clearly, study of a large sample of high redshift X-ray clusters is required in order for the evolution of the cluster scaling relations to be adequately constrained. 

%=============================================================================================
\acknowledgments

We thank the anonymous referee for some helpful comments which helped to improve the quality of this paper. This work is based on data obtained by \textit{XMM-Newton}, an ESA science mission funded by contributions from ESA member states and from NASA. We acknowledge financial support from the NASA-LTSA program, the RAS Hosie Bequest, Liverpool John Moores University, the Institute for Astronomy at the University of Edinburgh, the XMM and Chandra guest observer programs, Carnegie Mellon University, the NSF, and PPARC. This work was performed under the auspices of the U.S. Department of Energy, National Nuclear Security Administration by the University of California, Lawrence Livermore National Laboratory under contract No. W-7405-Eng-48. This research made use of the NASA/IPAC Extragalactic Database, the SIMBAD facility at CDS, the NASA/GSFC-supported XSPEC software, and the ESO Imaging Survey. Based in part on observations obtained at the European Southern Observatory (ESO Programme 078.A-0060). The W. M. Keck Observatory is a scientific partnership between the University of California and the California Institute of Technology, made possible by a generous gift of the W. M. Keck Foundation. The authors wish to recognize and acknowledge the very significant cultural role and reverence that the summit of Mauna Kea has always had within the indigenous Hawaiian community; we are fortunate to have the opportunity to conduct observations from this mountain. The analysis pipeline used to reduce the DEIMOS data was developed at UC Berkeley with support from NSF grant AST 00-71048.

{\it Facilities:} \facility{Keck:II}, \facility{VLT:Antu}, \facility{XMM}.

%=============================================================================================

\clearpage

\begin{deluxetable}{cccccc}
\tablewidth{0pt}
\tabletypesize{\scriptsize}
\tablecaption{Spectroscopic observations log.\label{t_spectroObsData}}
\tablehead{\colhead{Mask}	& \colhead{Slits}	& \colhead{Grating+Filter} & \colhead{Frames} & \colhead{Airmass}	& \colhead{Date (UT)}}
\startdata
\sidehead{Keck II (DEIMOS):}
1		& 81			& 600ZD+OG550	& $5\times 1800$ sec	& 1.5		& 16/09/2006\\
2		& 143		& 600ZD+OG550	& $8\times 1800$ sec	& 1.3		& 20/09/2006\\
3		& 141		& 600ZD+OG550	& $7\times 1800$ sec	& 1.3		& 21/09/2006\\
\sidehead{Antu (FORS2):}
1               & 11                    & 300I+OG590 & $9\times 900$ sec    & 1.2 & 04/07/2006   \\
2               & 12                    & 300I+OG590 & $6\times 900$ sec    & 1.2 & 20/10/2006   \\
3               & 12                    & 300I+OG590 & $3\times 900$ sec    & 1.4 & 21/10/2006   \\
4               & 12                    & 300I+OG590 & $9\times 900$ sec    & 1.1 & 15-18/11/2006\\
\enddata
\end{deluxetable}

% fudge to align on decimal point in table
\newdimen\digitwidth
\setbox0=\hbox{\rm0}
\digitwidth=\wd0
\catcode`?=\active
\def?{\kern\digitwidth}

\begin{deluxetable}{ccccccccc}
\tablewidth{0pt}
\tabletypesize{\scriptsize}
\tablecaption{Spectroscopic members of the cluster J2215.9$-$1738.\label{t_J22159members}}
\tablehead{\colhead{Object ID}	& \colhead{R.A. (J2000)}		& \colhead{Dec. (J2000)}		& \colhead{$z$}	& \colhead{$Q$\tablenotemark{a}} & \colhead{Method\tablenotemark{b}} & \colhead{Telescope}	& \colhead{\citet{Stanford_2006}}}
\startdata
\sidehead{Objects within the virial radius ($R_{\rm v} = 1.05$ Mpc):}
1	& 22:15:58.478	& $-$17:37:58.58		& \phantom{0}1.452$\pm$0.001\phantom{0}	& 3	& V	& Keck	& Object 14389\\
2	& 22:15:58.905	& $-$17:37:59.12		& \phantom{0}1.451$\pm$0.001\phantom{0}	& 3	& V	& Keck	& Object 14378\\
3	& 22:15:59.035	& $-$17:38:02.50		& \phantom{0}1.454$\pm$0.001\phantom{0}	& 2	& V	& VLT	& Object 14339\\
4	& 22:15:58.480	& $-$17:38:10.71		& 1.4650$\pm$0.0003						& 3	& X	& VLT	& Object 14289\\
5	& 22:15:59.080	& $-$17:38:02.40		& \phantom{0}1.459$\pm$0.001\phantom{0}	& 2	& V	& VLT	& \nodata\\
6	& 22:15:58.380	& $-$17:38:10.71		& \phantom{0}1.466$\pm$0.001\phantom{0}	& 1	& V	& VLT	& \nodata\\
7	& 22:15:58.850	& $-$17:38:10.89		& \phantom{0}1.453$\pm$0.001\phantom{0}	& 2	& V	& VLT	& \nodata\\
8	& 22:15:59.174	& $-$17:37:53.94		& 1.4619$\pm$0.0003						& 3	& X	& Keck	& \nodata\\
9	& 22:15:57.441	& $-$17:37:57.86		& \phantom{0}1.454$\pm$0.001\phantom{0}	& 2	& V	& VLT	& \nodata\\
10	& 22:15:59.707	& $-$17:37:59.16		& \phantom{0}1.469$\pm$0.001\phantom{0}	& 1	& V	& VLT	& \nodata\\	
11	& 22:15:57.220	& $-$17:38:07.80		& 1.4502$\pm$0.0001						& 3	& X	& Keck	& \nodata\\
12	& 22:15:59.870	& $-$17:37:59.23		& \phantom{0}1.449$\pm$0.002\phantom{0}	& 1	& V	& VLT	& \nodata\\
13	& 22:15:57.232	& $-$17:37:53.11		& 1.4537$\pm$0.0001						& 3	& X	& Keck	& Object 14478\\
14	& 22:15:58.363	& $-$17:37:37.48		& 1.4526$\pm$0.0001						& 3	& X	& Keck	& Object 14651\\
15	& 22:16:00.705	& $-$17:37:51.02		& \phantom{0}1.471$\pm$0.001\phantom{0}	& 2	& V	& Keck	& \nodata\\
16	& 22:15:56.186	& $-$17:37:49.83		& 1.4545$\pm$0.0003						& 3	& X	& VLT	& \nodata\\
17	& 22:15:56.059	& $-$17:37:49.90		& \phantom{0}1.461$\pm$0.001\phantom{0}	& 2	& V	& VLT	& \nodata\\
18	& 22:15:59.448	& $-$17:38:38.18		& 1.4569$\pm$0.0001						& 3	& X	& Keck	& \nodata\\
19	& 22:15:56.316	& $-$17:37:37.95		& \phantom{0}1.449$\pm$0.001\phantom{0}	& 1	& V	& VLT	& \nodata\\
20	& 22:16:03.096	& $-$17:38:08.05		& \phantom{0}1.462$\pm$0.001\phantom{0}	& 1	& V	& VLT	& \nodata\\
21	& 22:16:03.158	& $-$17:38:29.68		& 1.4650$\pm$0.0002						& 1	& X	& VLT	& \nodata\\
22	& 22:15:52.483	& $-$17:37:46.16		& \phantom{0}1.461$\pm$0.001\phantom{0}	& 2	& V	& VLT	& \nodata\\
23	& 22:15:51.664	& $-$17:37:11.96		& 1.4612$\pm$0.0004						& 3	& X	& VLT	& \nodata\\

 \sidehead{Additional objects within 2 Mpc:}
24	& 22:15:54.921	& $-$17:40:13.29		& 1.4752$\pm$0.0003						& 2	& X	& Keck	& \nodata\\
25	& 22:16:09.525	& $-$17:38:26.23		& 1.4613$\pm$0.0001						& 3	& X	& Keck	& \nodata\\
26	& 22:16:10.303	& $-$17:36:23.36		& 1.4632$\pm$0.0001						& 3	& X	& Keck	& \nodata\\
27	& 22:16:04.156	& $-$17:34:23.08		& 1.4737$\pm$0.0001						& 3	& X	& Keck	& \nodata\\
\enddata
\tablenotetext{a}{Redshifts with $Q=1$ are not secure and were not used in the analysis in \textsection~\ref{s_SpectroResults}.}
\tablenotetext{b}{X = cross-correlation, V = visual inspection (see \textsection~\ref{s_SpectroAnalysis} for details).}
\end{deluxetable}

\begin{deluxetable}{lcccl}
\tablewidth{0pt}
\tabletypesize{\scriptsize}
\tablecaption{Literature data used in Fig.~\ref{f_sigmaTnormalisation_evo}.\label{t_SigmaTLiterature}}
\tablehead{\colhead{Object}& \colhead{$z$}	& \colhead{$\sigma_v$\tablenotemark{a}} & \colhead{$T$\tablenotemark{b}}	& \colhead{Reference(s)}}
\startdata
RX J0848$+$4456		& 0.570	& \phantom{0}$670 \pm 50$\phantom{0}	& \phantom{0}$3.6 \pm0.4$		& \citet{Holden_2001}\\
RDCS J0910$+$5422	& 1.10\phantom{0} & \phantom{0}$675 \pm 190$	& \phantom{0}$7.2 \pm 1.8$		& \citet{Stanford_2002}, \citet{MeiRDCS_2006}\\
RX J0152.7$-$1357 S	& 0.833	& \phantom{0}$301 \pm 115$	& \phantom{0}$5.2 \pm 1.0$		& \citet{Girardi_2005}, \citet{MaughanJ0152_2006}\\
RX J0152.7$-$1357 N	& 0.833	& \phantom{0}$888 \pm 110$	& \phantom{0}$5.5 \pm 0.9$		& \citet{Girardi_2005}, \citet{MaughanJ0152_2006}\\
MS 0016$+$1609		& 0.541	& $1127 \pm 140$	& \phantom{0}$9.9 \pm 0.5$		& \citet{Vikhlinin_2002}, \citet{Borgani_1999}\\
MS 0451$-$0305		& 0.537	& $1330 \pm 100$	& \phantom{0}$8.1 \pm 0.8$		& \citet{Vikhlinin_2002}, \citet{Borgani_1999}\\
Cl J1226.9$+$3332		& 0.892	& \phantom{0}$997 \pm 250$	& $11.5 \pm 1.0$		& \citet{Maughan_2004}\\
RDCS J1252.9$-$2927	& 1.237	& \phantom{0}$747 \pm 80$\phantom{0}	& \phantom{0}$6.0 \pm 0.6$		& \citet{Rosati_2004}, \citet{Demarco_2007}\\ 
\enddata
\tablenotetext{a}{km s$^{-1}$}
\tablenotetext{b}{keV}
\end{deluxetable}

%=============================================================================================
\clearpage

%=============================================================================================

\end{document}